\documentstyle[12pt,epsf]{article}
\newlength{\abstractwidth}
\renewcommand{\thefootnote}{\fnsymbol{footnote}}
\renewcommand{\thanks}[1]{\footnote{#1}} % Use this for footnotes
\newcommand{\starttext}{
\setcounter{footnote}{0}
\renewcommand{\thefootnote}{\arabic{footnote}}}
\renewcommand{\theequation}{\thesection.\arabic{equation}}
\newcommand{\be}{\begin{equation}}
\newcommand{\bea}{\begin{eqnarray}}
\newcommand{\eea}{\end{eqnarray}}
\newcommand{\beq}{\begin{equation}}
\newcommand{\ee}{\end{equation}}
\newcommand{\eeq}{\end{equation}}

\def\ba{\begin{eqnarray}}
\def\ea{\end{eqnarray}}

\def\14{{1\over4}}
\def\12{{1 \over 2}}
\def\ap{area preserving diffeomorphism}
\def\aps{area preserving diffeomorphisms}
\def\ep{\epsilon_{ij}}
\def\cs{Chern Simons}
\def\qp{quasiparticle}
\def\qps{quasiparticles}
\def\qh{Quantum Hall }
\def\ro{\rho_0 }
\def\epa{\epsilon_{ab} }
\def\nc{non-commutative }
\def\ncy{non-commutativity }

\def\f{{\cal F }}
\def\ah{\hat{A}}
\def\fh{\hat{F}}
\def\eq{&=&}
\def\adag{a^{\dag}}
\def\bdag{b^{\dag}}
\begin{document}
\renewcommand{\theequation}{\thesection.\arabic{equation}}
\begin{titlepage}
\bigskip
\rightline{SU-ITP 00-25}
\rightline{hep-th/0101029}

\bigskip\bigskip\bigskip\bigskip

\centerline{\Large \bf {The Quantum Hall Fluid and }}
\centerline{\Large \bf { Non--Commutative
Chern Simons Theory}}

\bigskip\bigskip
\bigskip\bigskip
%\centerline{\it }
%\medskip
%\centerline{} \centerline{} \centerline{}
%\bigskip

\centerline{\it Leonard Susskind }
\medskip
\centerline{Department of Physics} \centerline{Stanford
University} \centerline{Stanford, CA 94305-4060}
\medskip
\medskip

\bigskip\bigskip
\begin{abstract}
The first part of this paper is a review of the author's
work with S. Bahcall which
gave an elementary derivation of the Chern Simons description of
the Quantum Hall effect for filling fraction $1/n$.
The notation has been modernized to
conform with standard gauge theory conventions.

In the second part arguments are given to support the claim that
abelian non--commutative Chern Simons theory at level $n$ is
exactly
equivalent to the Laughlin theory at filling fraction $1/n$. The
theory may also be formulated as a matrix theory similar to that
describing D0--branes in string theory. Finally it can
also be thought of as the quantum theory of mappings between two
non--commutative spaces, the first being the target space and the
second being the base space.

\medskip
\noindent
\end{abstract}

\end{titlepage}
\starttext \baselineskip=18pt \setcounter{footnote}{0}

%%%%%%%%%%%%%%%%%%%%%%%%%%%%%%%%%%%%%%%%%%%%%%%%%%%%%%%%%%%%%%%%%%%%%%
%%%%%%
%%%%%%%%%%%%%%%%
%%%%%%%%%%%%%%%%%%%%%%%%%%%%%%%%%%%%%%%%%%%%%%%%%%%%%%%%%%%%%%%%%%%%%%
%%%%%%
%%%%%%%%%%%%%%%%
\setcounter{equation}{0}
\section{   Fluid Dynamics in Co-moving Coordinates  }

The configuration space of charged particles in a strong magnetic
field \footnote{A  magnetic field is considered strong if the energy
scales of interest are too low for higher Landau levels
to be excited or admixed into the wave function. } is a
\nc \ space. The charged
particles  behave unlike conventional relativistic or
non--relativistic particles in so much as they are locked in place
by the large magnetic field. By contrast, a neutral system such as
a dipole \cite{dipole, bigatti} does move like a
conventional particle although it grows
in size with its momentum. Such dipoles are the objects described
by \nc \ field theory.

If we are considering a system of charged particles in a strong
magnetic field  we may either describe the  system in terms of
charge carrying fields such as the electron field or in terms of
neutral fields such as the density and current. The former fields
correspond to the frozen particles but the latter fields carry the
quantum numbers of dipoles. This suggests that the currents and
density of a system of electrons in a strong magnetic field may be
described by a \nc \ quantum field theory.
In this paper we will show that this is indeed the case and that
the Laughlin electron theory of the fractional quantum hall states
is equivalent to a \nc \ \cs \ theory describing the density and
current.

The first six sections   of the paper
review work done  in 1991 with
Safi Bahcall \cite{safi}.
The  purpose is to give an elementary  description of the
fractional \qh \ fluid for the case in which the inverse filling
fraction
is  an integer, and to explain how in these cases, the long distance
behavior of the \qh \ fluid can be described by \cs \ theory \cite{scz}.

Both the odd and even integer cases describe quantum hall states,
the odd cases corresponding to fermions and the even to bosons.

In the remaining sections it is shown that precise quantitative
agreement with Laughlin's  theory \cite{bob} can be obtained if the
ordinary
\cs \ theory is replaced by the \nc \ theory \cite{connes}.
Alternatively
it can be formulated as a matrix theory similar to that describing
D0--branes in string theory \cite{bfss}.

We will begin with a description of a dissipationless
fluid.
Consider a collection of identical
non--relativistic particles, indexed by $\alpha$,  moving
on a plane with Lagrangian
\be
L=\sum_{\alpha}{m\over 2}\dot{x}_{\alpha}^2 -U(x)
\ee
where $U$ is the potential energy.  Assuming the system
behaves like a fluid we can pass to a continuum
description by replacing the discrete label $\alpha$ by a pair of
continuous coordinates
$y_1,y_2$. These coordinates label the material points of the
fluid and move with it \footnote{Fluid mechanics described
in the $y$ coordinates is called the Lagrangian description. The
Eulerian
description expresses the fluid properties as functions of $x$.}. They
are the analog of co-moving
coordinates in cosmology. The system of particles is thereby
replaced by a pair of continuum fields $x_i(y,t)$ with $i=1,2$.
Without loss of generality we can choose the coordinates $y$ so
that the number of particles per unit area in $y$
space is constant and given by
$\ro$ . The real space density is
\be
\rho = \ro \left|  {\partial y \over \partial x}\right|
\ee
where
$ \left|  {\partial y \over \partial x}\right|$ is the Jacobian
connecting the $x$ and $y$ coordinate systems.

The potential $U$ is assumed to arise out of short range forces
which  lead to an equilibrium when the real space density
is $\ro$.
Thus in equilibrium the Jacobian is $1$. With these
assumptions and
conventions the Lagrangian can written as
\be
L=\int d^2 y \ro
\left[  {m\over 2} \dot{x}^2 - V\left(\ro
\left|  {\partial y \over \partial
x}\right|\right)                                              \right]
\ee
where the potential energy has now been expressed in terms of the
density \footnote{We are considering the case
of zero temperature. More generally the potential can also be
a function of the temperature. However, in order to study sound waves
with
small amplitudes we also should impose the "adiabatic condition". This
will lead to a potential which is $\rho$ dependent. I am grateful to
M.M. Sheikh
Jabbari for pointing this out.}.

The Lagrangian (1.3) has an exact gauge invariance under \aps \ of
the $y$ plane.
Consider any \ap \ from $y$ to $y'$ with unit Jacobian. The fluid
field $x$ transforms as a scalar, $x'(y') =x(y)$. It is easily
seen that (1.3) is invariant. To find the consequences of this
invariance consider an infinitesimal transformation
\be
y'_i =y_i + f_i(y).
\ee
The $x's$ transform as
\be
\delta x_a= {\partial x_a \over  \partial y_i} f_i(y).
\ee

The condition for $f$ to represent an infinitesimal \ap \ is that
\be
f_i = \ep {\partial \Lambda(y) \over \partial y_j}
\ee
with $\Lambda$ being an arbitrary gauge function. Equation (1.5)
then takes the form
\be
\delta x_a= \ep {\partial x_a\over\partial y_i}
{\partial \Lambda \over \partial y_j}
\ee

\setcounter{equation}{0}
\section{ Kelvin's Circulation Theorem and Vortices }

Since the transformation (1.7) is a symmetry of $L$, a conserved
quantity exists and is given by
\be
\int d^2 y \Pi_a \delta x_a
\ee
where $\Pi_a$ is the canonical conjugate to $x_a$, proportional to
${\dot{x}}_a$. Thus for any $\Lambda$
\be
\int d^2y \ro \left[
\ep \dot{x}_a
{\partial x_a\over \partial y_i }{\partial \Lambda \over\partial y_j}
\right]
\ee
is conserved. Integrating by parts gives
\be
{d \over dt} \int d^2y \ \ \ep {\partial \over \partial y_j}
\left[\dot{x}_a      {\partial x_a \over \partial y_i}  \right]
\Lambda =0.
\ee
Since eq(2.3) is true for all $\Lambda$ we can conclude that
\be
{d \over dt}\left[{\partial \over \partial y_j}
\left(    \ep \dot{x}_a  { \partial x_a \over \partial y_i}
 \right)
\right]
=0
\ee

To see how Kelvin's circulation theorem comes about, integrate
(2.4) over an area bounded by a closed curve $\Gamma$. Using
Gausse's theorem we find
\be
{d \over dt} \oint_{\Gamma} \dot{x}_a dx_a =0.
\ee
Thus conservation of circulation follows from gauge invariance in
the same way that
Gauss' law
is derived in free electrodynamics.

In free electrodynamics points where $\nabla \cdot E $ is not
zero correspond to static charges. In a similar way points where
${\partial \over \partial y_j}
\left(    \ep \dot{x}_a  { \partial x_a \over \partial y_i}
 \right) \neq 0
$
are vortices which are frozen into the fluid. The vortex--free
fluid satisfies
\be
\left[{\partial \over \partial y_j}
\left(    \ep \dot{x}_a  { \partial x_a \over \partial y_i}
 \right)
\right]
=0
\ee

\setcounter{equation}{0}
\section{ Electromagnetic Analog  }

The analogy with electromagnetic theory can be made much closer by
restricting attention to small motions of the fluid. Assuming the
potential $V$ in (1.3) has a minimum at $\rho = \ro$, there is a
time independent
solution of the equations of motion given by
\be
x_i=y_i.
\ee
Now consider small deviations from this equilibrium solution
parameterized by a vector field $A$ defined by
\be
x_i = y_i +\ep {A_j \over 2\pi \ro}
\ee
Working to linear order the gauge transformation (1.7) becomes
\be
\delta A_i =  2\pi \ro {\partial \Lambda \over \partial y_i}
\ee
which has the standard form of an abelian gauge transformation.
The exact form of the transformation is
\be
\delta A_i =  2\pi \ro {\partial \Lambda \over \partial y_i}
+{\partial A_i \over \partial y_l}{\partial \Lambda \over \partial y_m}
\epsilon_{l,m}.
\ee
The second nonlinear term in (3.4) is
suggestive of a non-commutative structure for the field theory
describing the \qh \ fluid. We will return to this in section (7).

We assume that for small deviations of the density from its
equilibrium value the potential has the form
\be
V=\mu \left(
\ro \left|   {\partial y \over \partial x}  \right| - \ro
\right)^2.
\ee
The density to lowest order is
\be
\rho = \ro - {1\over 2\pi} (\nabla \times A)
\ee
and the Lagrangian takes the form
\be
L= {1\over g^2} \int d^2 y {1\over 2} \left[
\dot{A}^2 - {2\mu \ro^2 \over m}  (\nabla \times A)^2
\right].
\ee
where the coupling constant $g$ is defined by
\be
g^2 =(2\pi)^2 {\ro \over m}
\ee

The Lagrangian (3.7) is the familiar Maxwell Lagrangian in
temporal gauge. The velocity of light is given by
\be
c^2={2 \mu \ro^2 \over m}.
\ee
The photons of the analog electrodynamics are just sound waves in
the fluid.

The vortex free condition (2.6) in the linearized approximation
becomes the Gauss law constraint
\be
\nabla \cdot E = \nabla \cdot \dot{A} =0.
\ee
Thus we see that charges in the gauge theory represent vortices.

The equations of motion and the Gauss law constraint can be
derived from a single action principle by introducing a time
component for the vector field $A$. The procedure is well known
and will not be repeated here.

The derivation of the gauge description of fluid mechanics given
in this section was done in the temporal gauge $A_0 = 0$. However
once we have introduced $A_0$ back into the equations we are free
to work in other gauges. To some degree this freedom allows us to
relax the condition that the motion of the fluid be a small
deviation from the configuration $x=y$. In fact we may use the
gauge freedom to work in a gauge in which the displacement of the
fluid is as small as possible. To carry this out let us introduce
a positive measure $M$, for the magnitude of $A$.
\be
M=\int d^2y A(y)\cdot A(y).
\ee
We may choose our gauge by requiring that $M$ be as small as
possible, that is
\be
\delta M =0
\ee
where the variation is with respect to an arbitrary gauge
transformation  $\delta A = \nabla \lambda$. Thus for a given
configuration $A$ should be chosen to satisfy
\be
\int d^2y A\cdot \nabla \lambda =0.
\ee
This obviously requires the Coulomb gauge.
\be
\nabla \cdot A=0 \ee Thus by working in the Coulomb gauge we are
also insuring that the $y$ and $x$ coordinates agree as closely as
possible. More generally when the non--linearity and \ncy \ of the
equations is included we will define a generalization of the
Coulomb gauge which minimizes
\be
M= \int d^2y (x-y)^2.
\ee

\setcounter{equation}{0}
\section{ Charged Fluid in A Magnetic Field  }

Now let us assume that the particles making up the fluid are
electrically charged and move in a background magnetic field $B$.
For a point particle of charge $e$ in a uniform magnetic field the
Lagrangian gets an extra term
\be
{eB\over 2} \epa \dot{x}_a x_b
\ee
For a fluid with charge to mass ratio $e/m$ the extra term is
\be
L'= {eB \over 2} \int \ro d^2y \epa \dot{x}_a x_b.
\ee
Note that the canonical momentum density conjugate to $x_a$ is
given by
\be
\Pi_a ={\partial L' \over \partial {\dot{x}_a}}=   {eB\ro \over 2} \epa
x_b
\ee

Substituting (3.2) into (4.2) and dropping total time derivatives gives
\be
L'= {eB \over 8\pi^2 \ro} \int d^2y \epa \dot{A}_a A_b.
\ee
This has the usual form of an abelian \cs \ Lagrangian in the
temporal gauge. Among its effects are to give the photon a mass.
The mass is given by
\be
m_{photon} = eB/
m
\ee
which will be recognized as the cyclotron frequency.

In the absence of a magnetic field, the role of static charges was
to represent the fluid vortices. When the $B$ field is turned on
the long range behavior of the theory is dominated by the \cs
\ term and the character of the charges changes. In what follows
we will be mainly interested in the long distance behavior of the
\qh \ effect. In this case we may drop the Maxwell term
completely. Let us do so.

The Lagrangian (4.2) is  invariant under \aps . Accordingly
(2.1) is still conserved, but now the canonical momentum conjugate
to $x_a$ is $\Pi_a \propto \epa x_b$. The conserved gauge generator is
$$
\12 {\partial \over \partial y_j}\left\{\ep \epa x_b {\partial x_a
\over \partial y_i}
\right\} =
\12 \ep \epa {\partial x_b \over \partial y_j}
{\partial x_a \over \partial y_i}
$$
This is just the Jacobian from $x$ to $y$ which is given by $\ro /
\rho$. It therefore follows that the density of the fluid at a
fixed co-moving point $y$ is time independent.

In the absence of vortices (quasiparticles in the Quantum Hall
context) the conserved generator is set to unity. Thus the equations of
motion are supplemented with the constraint
\be
\12 \ep \epa {\partial x_b \over \partial y_j}
{\partial x_a \over \partial y_i}=1
\ee

The equations of motion and constraint can be obtained from a
single action by introducing a time component of $A$ and replacing
the ordinary time derivative in (4.2) by an appropriate covariant
derivative:
\be
L'={eB \ro \over 2} \epa \int d^2y   \left[
 \left(\dot{x_a} - {1\over 2 \pi \ro}\{x_a,A_0
\}\right)x_b +{\epa \over 2\pi \ro}A_0 \right ].
\ee
In this equation we have introduced the Poisson bracket notation
$$
\{ F(y),G(y) \} =  \ep \partial_iF \partial_jG
$$

Now return to the linearized approximation for small oscillations
of the fluid. Using the expression (3.6) for the density we see
that the conservation law requires the ``magnetic field"
\footnote{Warning:
Do not confuse the analog magnetic field $\nabla \times A$ with the
external magnetic field $B$ } at each point $y$, to be time
independent. The analog of a  vortex is a $\delta$ function magnetic
field:
\be
\nabla \times A = 2\pi \ro q \delta^2(y)
\ee
where $q$ measures the strength of the vortex
\footnote{Again we warn the reader not to confuse quantities
in the analog gauge theory with real electromagnetic quantities.
The real electric charge of an electron is $e$ and the analog
gauge charge of the vortex is $q$. We will see that the \qp \ also
carries a real electric charge}.
The solution to this equation is unique up to a gauge
transformation. In the Coulomb gauge, $\nabla \cdot A =0$, it is
given by
\be
A_i = {q \ro } \ep {y_j \over y^2}.
\ee

Since $\ep A_j/2\pi \ro$ is the displacement of the fluid we see that
the \cs \ vortex is really a radial displacement of the fluid
toward or away from the vortex-center by an amount $q/2 \pi r$
depending on the sign of $q$.
This implies either an excess or deficit of ordinary
electric charge at the vortex.
The magnitude of this excess/deficit is
\be
e_{qp} = \ro q e.
\ee
The charged vortex is the Laughlin \qp  \cite{bob}.

To further understand the \qp \ we must quantize the fluid. We
will not carry out a full quantization but instead rely on
elementary semiclassical methods. Assume the fluid is composed of
particles of  charge $e$. If $\Pi_a$ is the momentum
density then
\be
p_a = {\Pi_a \over \ro  } =eB \epa x_b/2
\ee
is the momentum of a single particle. The standard Bohr--Sommerfeld
quantization condition is
\be
\oint p_a dx_a =  2\pi n.
\ee
The quantization
condition (4.12) becomes
\be
{eB } \oint { \epa x_b \over 2} dx_a  = 2\pi n .
\ee
The integral in (4.13) is the real ($x-space$) area of the region.
To interpret the meaning of this equation we shall assume that any
change in the properties of the fluid within the closed curve,
such as the introduction of a \qp \ can only change $eB \times (area)$
by $2\pi$ times an integer. Thus
\be
{eB \over 2\pi \ro}\oint A_a dy_a = 2 \pi n .
\ee
Using the vortex solution (4.9) then gives
\be
eBq=2\pi n
\ee
{}From (4.10) a single elementary \qp ($n=1$) has electric charge
\be
e_{qp} = 2\pi{ \ro  \over B}
\ee
This agrees with the value of the quasiparticle charge from
Laughlin's theory \cite{bob}.

According to (4.9) the vector potential diverges at the vortex. To
correctly understand the physics very close to the origin we must
give up the approximation of small disturbances. Doing so we will
see that the solution is well behaved. The correct equation for
the vortex is obtained by modifying  (4.6) to include a source,
\be
\12 \ep \epa {\partial x_b \over \partial y_j}
{\partial x_a \over \partial y_i} -1 = q \delta^2(y).
\ee
This equation has the solution
\be
x_i=y_i \sqrt{1+{q\over \pi |y|^2}}.
\ee
Far from the origin the solution agrees with (4.8) but has a
more interesting behavior near $y=0$. Although the vortex is a
point in $y$ space it has finite area in $x$ space. The leading
behavior is given by
\be
x_i \sim \sqrt{q \over \pi} {y_i \over |y|}.
\ee
The point $y=0$ is mapped to a circle of radius $\sqrt{q/\pi}$,
leaving an empty hole in the center. The hole has area $q$  and an
electric charge deficit $\ro q e/m$.

Before continuing we will introduce some notation which will be
helpful in relating the parameters $eB$ and $\ro$ with
field theoretic parameters. The quantity
\be
\nu  \equiv {2 \pi \ro \over eB}
\ee
is the ratio of the number of electrons to the magnetic flux and
is called the filling fraction. Comparing (4.4) with the
conventional \cs \ notation we see that $1/{\nu}$ is the \cs \
$level$ usually called $k$. In terms of the filling fraction, the
quasiparticle charge (4.16) is
\be
e_{qp} = e \nu
\ee

The parameter
\be
\theta \equiv {1 \over 2 \pi \ro}
\ee
appearing in (3.2) will later be identified with the
non-commutativity parameter of the non-commutative
coordinates that replace the y's in
section (6).

\setcounter{equation}{0}
\section{ Fractional Statistics of Quasiparticles }

In this section we will give an elementary explanation of why \qps
\ have fractional statistics \cite{bob}. The fractional statistics
question
can be reduced to the calculation of the Berry phase induced by
transporting one \qp \ around another. The calculation of this
phase depends only on the fact that when a \qp \ is created in the
fluid it pushes the fluid out by a distance $q /2 \pi r$ where $r$
is the distance from the \qp .

In the undisturbed fluid a \qp \ at the origin can be created by a
unitary operator having
the form
\be
U(0) = \exp{ \left \{
{iq\over 2 \pi}\int d^2y{\Pi(r) \cdot r \over  r^2} \right \}}
\ee
where $\Pi$ is the canonical conjugate to $x$ and $r$ is the
distance from the origin. A similar operator $U(a)$ can be
constructed which creates a \qp \ at $y=a$. It is important to
remember that the operator $U(0)$ not only creates a \qp \ at $y=0$
but also pushes the fluid away by distance $q/2 \pi |r|$.

Now let us construct a pair of \qps \, one at point $a$ and one
at  $b$. The naive guess would be
\be
| a,b\rangle = U(a) U(b)|0\rangle.
\ee
However this is not right. The first operator to act, $U(b)$,
creates a quasiparticle at $x=y =b$. Then $U(a) $ acts to create
a \qp \ at $x=y =a$ but it also pushes the fluid so that the first
\qp \ ends up at a shifted position. The right way to create the
\qps \ is to compensate for this effect by shifting the argument
of the first operator;
\be
| a,b\rangle = U(a) U(b - d_{a,b})|0\rangle.
\ee
where
\be
d_{a,b} = {q\over 2 \pi} {{a-b} \over |a-b|^2}.
\ee
This time  the creation of
the \qp \ at $a$ pushes the center of the first \qp \ to its
correct location at $x=b$.

Let us consider the Berry phase picked up by the wave function
when the \qp \ at $b$ is transported around a circle centered
at the fixed point $a$.
\be
\Gamma_{a,b} = \oint \left\langle
a,b \left|
{\partial \over \partial b}
\right|a,b\right\rangle
\ee
which can also be written
\be
\Gamma_{a,b} = \oint \left(
\langle a,b| a, b + db
\rangle
-1
\right).
\ee
The inner product $\langle a,b| a, b + db
\rangle$ is given by
\bea
&\langle 0|& U^{\dag}(b - d_{b,a})  U^{\dag}(a) U(a) U(b +db -d_{a,
b+db})
\ |0\rangle = \cr
&\langle 0|& U^{\dag}(b - d_{b,a})  U(b +db -d_{a, b+db})
 \ |0\rangle.
\eea
This last expression, when inserted into (5.6) gives the phase
for a state with only one \qp \ moved in a circle of radius
smaller by $|d_{ab}|$. Writing $R = |a-b| $ and $\Delta R = d_{a,b}$
we have
\be
\Gamma_{a,b}(loop \  with \ radius \ R)
=  \Gamma_b(loop \ with \ radius \ R-\Delta R).
\ee

Thus the extra phase due to the presence
of the \qp \ at $a$ is
\be
\Delta \Gamma = \Gamma_b(R) - \Gamma_b(R-\Delta R).
\ee

To compute $\Gamma_b(R)$ is easy because it is just the phase due
to moving a charge in a  uniform magnetic field. It is the
product of the charge times the enclosed flux. Since the charge of
the \qp \ is $Q = q \ro e $ the difference in phase is
\be
\Delta \Gamma = 2 \pi  R \Delta R Q B = \ro e B q^2.
\ee
Inserting the quantized value of $q$ from (4.14), $q= 2\pi  /
eB$ and  defining the filling fraction $\nu$
\be
\nu = {2 \pi \over eB}{\ro }
\ee
we find
\be
\Delta \Gamma = 2 \pi  \nu.
\ee
The parameter $\nu$ is the ratio of the particle number to the
magnetic flux and will be recognized as the usual filling
fraction. The connection between filling fraction and Berry phase
given in (5.12) is the same as derived by Laughlin and is
equivalent to the usual anyon statistics for the \qp .

\setcounter{equation}{0}
\section{Quantization of the Filling Fraction }

The simplest fractional \qh \ states are those for which the
filling fraction $\nu$ is of the form $1/n$ with $n$ being an
integer. Furthermore if the charged particles  comprising the
fluid are fermions (bosons)  then the integer $n$ should be odd
(even).  This  is a quantum mechanical effect related to
angular momentum quantization. We can very roughly
see how it comes about by
considering a pair of nearest neighbor particles in the fluid. The
relative angular momentum of the pair is
\be
L_{1,2} = \12 \epa (x_1 -x_2)_a (p_1-p_2)_b
\ee
Using $p_a = {eB\over 2} \epa x_b$ we find
\be
L_{1,2} = {1\over 4} eB \delta^2.
\ee
where $\delta$ is the separation between neighbors. In the ground state
it satisfies
\be
\delta \sim \sqrt{1/ \ro}
\ee
and
\be
L_{1,2} \sim eB/ \ro.
\ee
If we require this to be an odd (fermions) or even
(bosons) integer  we get
\be
eB/\ro \sim n
\ee
or
\be
\nu \sim 1/n.
\ee

It should be clear that this argument is at best a heuristic
suggestion of why we
may expect a quantization of the inverse filling fraction.
To find the precise quantization condition requires a more
exact quantization of the fluid such as that provided by the Laughlin
wave functions.
In the next section we will show that  exact agreement with the
Laughlin theory
 can be obtained by a
generalization of the fluid model in which ordinary abelian \cs \
theory is replaced by \cs \ theory on a \nc \ space \cite{connes}.
The main result is that we will rigorously  find the  connection
between filling fraction and particle statistics demanded by
Laughlin's theory.

\setcounter{equation}{0}
\section{Non--Commutative Chern Simons Theory  }

Let us consider the full nonlinear fluid equations which result
from the Lagrangian (4.7), the gauge invariance (3.4) and the
constraint (4.6). We introduce the Poisson bracket notation
\be
\{ F(y),G(y) \} =  \ep \partial_iF \partial_jG
\ee
The Lagrangian then takes the form \cite{rpm}
\be
L'= {1 \over 4\pi \nu} \epsilon_{\mu \nu \rho}\left[
{\partial A_{\mu} \over \partial y_{\rho}} -
{\theta \over 3}\{ A_{\mu},A_{\rho} \}
\right] A_{\nu}
\ee
where the indices $(\mu,\nu,\rho)$  run over $(0,1,2)$ and
$\theta = 1/2\pi \ro$.

Similarly  the gauge invariance and the constraint take the form
\be
\delta A_a= {\partial \lambda \over \partial y_a} +
\theta
\left\{ A_a,\lambda  \right\}
\ee
and
\be
\epsilon_{ab} \left[
\partial_a A_b -{\theta \over 2} \left\{ A_a,A_b  \right\}
\right] =0
\ee

The nonlinear theory defined by (7.2), (7.3) and ( 7.4)
is a Chern Simons gauge theory based on the group of area
preserving diffeomorphisms (APD's) of the parameter space $y_i$. In the
real electron system the $y$ space is just a convenient and
approximate way to label the electrons. Area preserving
transformations are merely a way of re--labeling
the particles. The correct way to label electrons is with a
discrete index and the re--labeling symmetry is the permutation
group. The gauge theory of APD's captures many of the long
distance features of the Quantum Hall system but it does not
capture the discrete or granular character of the electron system.

There is a well known way of discretizing the APD's. Two
equivalent lines of reasoning lead to the same conclusion. The
first is to recognize that (7.2), (7.3) and ( 7.4) are first order
truncations of a non--commutative theory, that is a field theory
on a \nc \ $y$--space. This theory is the \nc \ version of \cs \
theory with spatial \ncy . It is defined by the Lagrangian
\footnote{We have used the notation of Seiberg and Witten in which \nc \

gauge fields are ``hatted" }
\be
L_{NC}={1 \over 4 \pi \nu}
\epsilon_{\mu \nu \rho}
\left( \ah_{\mu}\ast \partial_{\nu} \ah_{\rho} +{2i \over 3}
\ah_{\mu} \ast  \ah_{\nu} \ast  \ah_{\rho}
\right)
\ee
with the usual Moyal star--product defined in terms of a \ncy \
parameter
\be
\theta = {1 \over 2\pi \ro}.
\ee
The Lagrangian (7.2) is identical to (7.5) expanded to first order
in $\theta$. On the other hand the full \nc \ theory is defined on
a \nc \ space in which there is a discrete indivisible unit of
$y$--space
area which can be identified with the electron.

A point worth discussing is which of the two sets of coordinates
$x$ or $y$ are \nc . The answer is both but in different senses.
The \nc \ \cs \ theory that we are considering is defined by
fields that are functions of the $y$ coordinates, that is the $y$ space
is
the base space.  The
Moyal brackets are defined in terms of $y$ derivatives. Evidently
it is the $y$ coordinates which are \nc \ with \ncy \ parameter
$\theta$. The basic quantum of $y$--area is $\theta
$ and it represents the area occupied by a single electron.
The \ncy \ of the $y$ space is classical and not due to the
quantization of the field theory.

On the other hand the Lagrangian (4.2) makes plain the fact that
the $x$ space is also \nc \ but in the quantum sense. From (4.11) we see

that the momentum
conjugate to the coordinates are proportional to the coordinates
themselves. Indeed the
coordinates $x_1,x_2$ do not commute as quantum objects.
In this case the \ncy \
parameter is $1/eB$ which is proportional to
the area occupied, not by an electron, but by a single quantum of
magnetic flux.
Evidently the \nc \ \cs \ theory describes mappings between two
\nc \ spaces.

Another route to the same theory which emphasizes the
discrete particle aspects of the fluid  begins with a matrix theory
representation of the electrons \footnote{The use of matrix theory
in this paper is different from that in \cite{t-nut, gubser}. In
that case the electrons were described by string ends and the
matrix theory described the units of magnetic flux or D0-branes.
In this paper
the electrons are described by matrix theory as if they were
D0-branes. The relation between the two descriptions will be
discussed in a forthcoming paper with N. Toumbas and B.
Freivogel.
} in a manner similar to the
construction of the matrix theory of D0--branes \cite{bfss}. We replace
the
classical
configuration space of $K$ electrons by a space of two
$K \times K$
hermitian matrices $X_a$. The time component of the vector potential is
also replaced by an hermitian matrix. We will eventually let $K$ be
infinite.
The natural action, generalizing
(4.7) to  matrix
theory in a background magnetic field is \cite{t-nut, kluson, poly}
\be
L'={eB\over  2}\epa Tr\left( \dot{x_a}-i [x_a, \ah_0]_m
\right)x_b +eB \theta \ah_0.
\ee
In this equation the notation $[f,g]_m$ indicates $f$ and $g$ are
classical matrices and the subscript $m$ means that the commutator
is evaluated in the classical matrix space and not in the Hilbert
space of quantum mechanics. Quantum commutators will be denoted in
the usual way with no subscript.

The equation of constraint is obtained by varying this action with
respect to $\ah_0$. We find
\be
[x_a,x_b]_m =i\theta \epa
\ee

It is well known that (7.8) can only be solved with infinite
matrices. Therefore we must allow the number of electrons $K$
to be infinite.

Now choose two definite matrices $y_a$ satisfying
\be
[y_a,y_b]_m =i\theta_{ab} = i \theta \epa
\ee
For example such matrices can be easily constructed from harmonic
oscillator creation and annihilation  operators. We can also
represent (7.9) in the form
$$
y_2 = -i\theta {\partial \over\partial y_1}
$$
Next define the
matrices $\ah_a$
\be
x_a= y_a + \epa \theta \ah_b.
\ee
Inserting (7.10) into (7.7) gives the \nc \ \cs \ Lagrangian (7.5).
Thus, from two points of view we see that \nc \ \cs \ theory
 is connected with the physics of charges moving in a magnetic
 field.

\setcounter{equation}{0}
\section{Statistics of the Chern Simons Particles }

If \nc \ \cs \ theory describes particles in a magnetic field,
what kind of particles are they? In particular are they fermions,
bosons, anyons or something new? The answer as we will see depends
on the level of the \cs \ theory  $1/\nu$.

The particles described by Matrix Theory \cite{bfss} satisfy a
more general statistics than either Fermi of Bose statistics. The
permutation of particle labels is replaced by the bigger group of
unitary transformations in the space of the matrix indices.
However certain backgrounds may break the unitary symmetry to the
subgroup of permutations. In that case the transformation property
under the subgroup will determine the statistics. In this section
we will compute the statistics of the particles defined by the
matrix theory of the previous section.

Let us begin with identification of canonical variables from the
action (7.7).  The
canonical momentum conjugate to matrix entry $(x_1)_{mn} $ is
\be
(p_1)_{nm}={eB \over 2}(x_2)_{nm}.
\ee
The $quantum$ commutation relations are not between $x_1$ and $x_2$
but between their matrix entries
\be
[(x_1)_{mn}, (x_2)_{rs}] ={2i\over eB}\delta_{ms} \delta_{nr}.
\ee
Another way to express this is
\be
(x_2)_{mn} = {-2i \over eB}{\partial \over \partial (x_1)_{nm}}.
\ee
Thus unlike the $y's$ which are classical non--commuting
coordinates, the matrix components of the $x's$ are \nc \ in the
quantum sense.

In order to simplify our notation we will work in the Hilbert
space basis in which $(x_1)_{mn} $ is diagonal. We will call
 $(x_1)_{mn} $ and $(x_2)_{mn} $ $X_{mn}$ and
${-2i \over eB}{\partial \over \partial X_{nm}}$ or more simply
 $(x_2)_{mn} =(2eB^{-1})P_{nm} $.

 We can now rewrite the constraint equation (7.8) in the form
\be
X_{mn}P_{nr}-P_{mn}X_{nr} =ieB \theta \delta_{mr}={i \over \nu}
\delta_{mr}
\ee
Since this equation was derived by varying with respect to $A_0$
it should be interpreted like the Gauss law constraint as acting
on a wave function whose arguments are $X_{mn}$.
\be
\left\{ X_{mn}P_{nr}-P_{mn}X_{nr}\right\}|\Psi \rangle =={i \over \nu}
\delta_{mr} |\Psi \rangle
\ee
This is the fundamental set of equations determining the ground
state of the \nc \ \cs \ theory.

The left hand side of (8.5) has a familiar form. It resembles an
angular momentum operator, that is a generator of rotations. In
fact it is the quantum mechanical generator that generates unitary
transformations among the matrix entries. For example consider an
Hermitian matrix $\lambda$ that generates infinitesimal
transformations according to
\be
\delta X = i [ X, \lambda]_m.
\ee
The corresponding quantum generator is
\be
\Lambda=  \lambda_{rm} \left\{ X_{mn}P_{nr}-P_{mn}X_{nr}\right\}
\ee
and the constraint (8.5) becomes
\be
\Lambda  |\Psi \rangle ={1\over \nu} Tr \lambda  |\Psi \rangle
\ee

Now let us turn to the question of the statistics of the charged
particles comprising the quantum hall fluid. Let us consider the
operation of exchanging two particles. Consider the case of just
two matrix theory particles described by $2 \times 2$ matrices.
The unitary matrix describing their interchange is obviously
\bea
U=\pmatrix{
0&1 \cr
1&0 \cr
}
\eea
More generally for $K \times K  $ matrices the exchange of the
$m^{th} $ and $n^{th}$ particle  can be
written as a matrix with two nonzero elements $U_{nm}=U_{nm}=1$,
and all other elements equal to zero. Furthermore we can also
write
\be
U=\exp{i \lambda}
\ee
such that $Tr \lambda = \pi.$
Equation (8.8) takes the form
\be
\Lambda  |\Psi \rangle ={\pi \over \nu}   |\Psi \rangle
\ee
Now consider  the exchange operation on the Hilbert
space of states. Call the unitary quantum  operator which implements
the exchange operation on the space of states ${\cal{P}}_{mn}$.
\be
{\cal{P}}_{mn} = exp{i \Lambda}.
\ee
Evidently
\bea
{\cal{P}}_{mn} |\Psi \rangle &=& exp{i \Lambda} |\Psi \rangle \cr
&=& \exp{\left(i \pi \over \nu \right)}  |\Psi \rangle.
\eea

This is a remarkable formula. It says that if the filling fraction
satisfies
$\nu = 1/(2n+1)$ the charged particles are fermions while if
$\nu = 1/(2n)$ they are bosons! This of course is identical to the
statistics--density connection implied by Laughlin's wave
function.

For more general values of $\nu$ the simple Laughlin wave
functions
\be
\Psi_{laugh} = \prod (Z_i-Z_j)^{1\over \nu} \exp{ \left(
-\12 \sum Z^\ast Z \right)}
\ee
should not be interpreted as quantum hall states for either
fundamental fermions or bosons. Their correct interpretation is as
wave functions for a system of anyons in a magnetic field. Thus we
expect that for non--integer $1/\nu$ the \nc \ cs \ theory has an
interpretation in terms of \qh \ states for fundamental charged
anyons. It is an interesting question how to describe the
non--integer filling of electrons by \nc \ field theory.

It is important to keep in mind that
quantities defined in the $y$ frame of reference are
not gauge invariant. Consider as an example the real space particle
density
defined in (3.6) and its non--linear generalizations. This
quantity is the density measured in $x$ coordinates but it is
naturally viewed as a function of $y$ in the formal field theory.
However functions of $y$ are not observables, indeed they are not
gauge invariant since $y$ itself changes under the area preserving
diffeomorphisms. The gauge invariant quantity is the density at  a
point in $x$ space. Thus
\be
\rho(x) =\rho(y+\epsilon \theta A)
\ee
is gauge invariant while $\rho(y)$ is not.
The density is closely related to the \nc \ field strength. In the
fluid equations of section (3) the density is given by
\be
\rho = \ro (1+\theta F)^{-1}
\ee
where
$$
F=\epa
\left(
{\partial A_b\over \partial y_a} +{\theta \over 2}\{ A_a, A_b
\} \right).
$$
In the more exact \nc \ theory the field strength $\f$ is given by
\be
\fh=\epa \left( {\partial \ah_b\over \partial y_a} +{1\over 2}
( \ah_a \ast \ah_b - \ah_b \ast \ah_a
) \right)
\ee
Since the physical density is given by (8.16) we should expect
that the value of the field strength at a location $x$ is a gauge
invariant
physical quantity. By contrast its value at a definite
value of $y$ is not. In fact it is well known that local
quantities in a \nc \ gauge theory are not gauge invariant.

Obviously the gauge invariant quantity can be obtained by a Taylor
series expansion in the $\theta$ parameter. Let us denote the value
of the field strength at the point $x$ by $f(x)$.
\be
f(x)=\fh(y_i+\ep \theta \ah_j)= \fh(y) +\ep \theta \ah_j\partial_i
\fh +....
\ee
This expansion is closely related to  the
Seiberg Witten \cite{seibwitt, seib}
map
\footnote
{This connection was explained to me by Nick Toumbas.}
 which  relates the gauge dependent \nc \ field
strength to a gauge invariant commutative field strength. The
Seiberg Witten map has the form
\be
F(x)=  \fh + \theta \fh^2  \fh(y) +\ep \theta \ah_j\partial_i
\fh +....
\ee
Where in this equation $F$ is a gauge invariant commutative field
strength. Evidently to the order we are working
\be
f(x) = F- \theta F^2.
\ee

The physical gauge invariant correlation functions are of the form
\be
\langle 0| f(x) f(x')       |0 \rangle
\ee
Calculating these correlation functions is obviously non--trivial.
The correlators of $\fh(y)$ are more straightforward. Carrying
out the Seiberg Witten map is difficult except
as a power series in $\theta$. This may be worth doing in order to
compare with Laughlin's theory which relates these correlation functions

to density--density correlations in a well defined Coulomb gas in
a neutralizing background. One interesting prediction is that as
the filling fraction decreases a phase transition occurs in which the
fractional
\qh \ states give way to a Wigner Crystal. This implies that the
\nc \ \cs \ theory also has a transition to a new phase at large
level.

\setcounter{equation}{0}
\section{The Phase Transition }

If there is a new phase at low filling fraction it is likely to be
associated with the breaking of a symmetry that we have not yet
discussed. To understand it let us return to the single particle
Lagrangian (4.1). This Lagrangian has a symmetry under area
preserving diffeomorphisms (APD's) of $x$ space. We emphasize that
the group of APD's of $x$ space is an entirely different symmetry
than the gauge symmetry of area preserving transformations of $y$
space.

 To see the symmetry under $x$--space APD's  consider the
infinitesimal APD
\bea
 x_i'&=&  x_i + \delta_i \cr
 \delta_i &=& \ep {\partial S(x) \over \partial x_j}
\eea where $S$ is a function of $x$. An easy calculation reveals
that the variation of the Lagrangian is a total time derivative:
\be
\delta L={d \over dt}\left[x_m{\partial S \over \partial x_m} -2S
\right]. \ee Hence the theory is invariant under APD of $x$ space.
This fact is unchanged when we pass to the  fluid Lagrangian
(4.2). The theory therefore exhibits  APD invariance of two kinds;
one which acts on the base coordinates  $y$, and the
other which acts on the target coordinates $x$.

When we pass to the quantum theory of a particle in a magnetic
field the APD are replaced by the corresponding transformations on
a \nc \ $x$ space. That the $x$ space is \nc \ is clear from the fact
that the two components $x_1,x_2$ are canonical conjugates of one
another. The quantum transformations which replace the classical APD's
are
the unitary transformations on the Hilbert space of LLL's.

As an example consider the quadratic functions
\be
S=c_{ij}x_ix_j
 \ee
with $c_{ij}$ being  traceless and symmetric.  In this case the total
time derivative vanishes and the Lagrangian is invariant. The APD
in this case are linear transformations. For the case $S=x_1 x_2  $
the finite transformations have the form
 \bea
  x_1'&=&cx_1 \cr
  x_2'&=&{1\over c}x_2.
 \eea
 That is they squeeze one direction and stretch the orthogonal
 direction.

We can construct the corresponding symmetries in the \nc \ quantum
theory most easily by focusing on the matrix version of the
theory. Consider the generator
 \be
 S=Tr c_{ij}x_ix_j.
 \ee
 Using the quantum commutation relations (8.2)
we find the matrix valued equation
 \bea
 \delta x_l=i[x_l,S]= {2\over eB}\epsilon_{jl}c_{ij}x_i
 \eea
 This is the matrix version of the
 APD generated by $S$.

 It is straightforward to show that $S$ commutes with the
 equations of constraint (8.4). Thus the Unitary transformation
 generated by $S$ is  a symmetry of the theory.
 It is therefore important to know how the symmetry is realized.
 We will argue that  the incompressible \qh \ fluid is
 invariant under transformations such as (9.4).

One reason for believing this is that the \qh \ fluid has a uniform
density which does not
 change under any APD. There is no obvious contradiction with
 saying the fluid is invariant.

 Quantum mechanically the transformations (9.4) induce a unitary
 transformation in the space of LLL's. For example the single
 particle wave function
$$
\psi = e^{-\12 Z^{\dag}Z}
$$
gets mapped to
$$
\psi' =e^{-\12 Z^{\dag}Z} e^{\12 \alpha Z^2}
$$
where $\alpha$ is a parameter representing the amount of squeezing
and stretching. By translating $\psi$ around on the
plane we can construct an (over) complete set of LLL's both before
squeezing or after.
It is not difficult to prove that the average
electron
occupation number for electrons in the original
states $\psi$ and the squeezed states $\psi'$ are the same.

For the case $\nu =1$ it is easy to see that the \qh \ state in
invariant under unitary transformations of the LLL's. This is
because the property having all fermion states filled is basis
independent.  A fully
filled system of fermion levels is fully filled in any basis. We
will assume without further proof that the all Laughlin states are
invariant under the unitary transformations of LLL's, at least for
the squeezing/stretching transformations.

As was mentioned, when the filling factor becomes sufficiently
small the system makes a transition to another phase, the Wigner
crystal phase. The transition is not driven by energetics. Indeed
the transition can be seen in the behavior of the Laughlin wave
functions themselves. While it may be correct that the relevance
of the Laughlin functions depends on the existence of repulsive
forces, the precise form of the wave functions corresponds to
vanishing potential.

In any case a crystal-like phase can not be invariant under
general APD's. It seems likely then that the crystal phase is
associated with a spontaneous breaking of the x-space APD's. It is
obvious that an APD acting on a crystal will change it and
take us to a new configuration. This in turn implies that the solutions
to the constraint equations will have to become degenerate.
It would be good to see this
directly from the equations of the \nc \ \cs \ theory
but at the moment it is a conjecture.

Another point may be very relevant in this context.  The existence
of the transition must be related to an instability of the
homogeneous fluid phase. Recent work has shown  a
generic tendency
for  \nc \ quantum field theories to exhibit  transitions to
 striped and other inhomogeneous phases.
\cite{stripegubser}.

 \setcounter{equation}{0}
\section{Non--Commutative Quasiparticles}

Let us consider the \nc \ generalization of (4.17) which defined a
\qp . The generalization of the left hand side of (4.17) is
obvious.
$$
 \ep  {\partial x_b \over \partial y_j}
{\partial x_a \over \partial y_i}  \to \theta [x_b, x_a ]_m.
$$
The more interesting question is how to represent the delta
function $\delta^2(y)$ on the \nc \ space. The correct answer is
that a delta function should be replaced by a projection operator
onto a particular vector in the matrix space \cite{solitons}. In order
to carry
this out in detail let  us introduce a particular description of
the infinite dimensional matrices. We begin by labeling basis
vectors \footnote{The notation $|n)$ will used for vectors in the matrix

space. For vectors in the quantum space of states we use $|\Psi\rangle
$.}
$|m)$
where $m$ runs over the positive integers and zero.
We also introduce matrices $a$ and $\adag$ with the usual
properties
\bea
\adag \  |n) \eq \sqrt{n+1} \ |n+1)\cr
a \ |n) \eq \sqrt{n} \ |n-1)
\eea

Equation (7.9) is   satisfied by expressing the
$y's$in terms of Fock space matrices;
\bea
y_1 \eq  \sqrt {\theta \over 2}  (a+a^{\dag}) \cr
y_2 \eq  \sqrt {\theta \over 2}  i(a-a^{\dag}).
\eea

A delta function at the origin may be represented as a projection
operator onto the vector $|0)$
\be
\theta \delta (y) \to |0)(0|.
\ee
The constraint equation becomes
\be
\theta^{-1}[x_1,x_2]_m = i + i \nu  \ |0)(0|.
\ee
This equation is equivalent to the
classical \nc \ field equation for
the quasiparticle.  The analysis of this equation
has similarities with that in \cite{solitons} where classical
soliton solutions of \nc \ field theory were found.

It is not difficult to solve (10.3)
exactly. First note that if we drop the
\qp \ term the constraint is solved by
\bea
x_1 \eq y_1=  \sqrt {\theta \over 2}  (a+a^{\dag}) \cr
x_2 \eq y_2 =  \sqrt {\theta \over 2}  i(a-a^{\dag}).
\eea

We can introduce two new matrices $b, b^{\dag}$ defined by the
action
\bea
\bdag \ |n) \eq \sqrt{n+1 +\nu} \ \ |n+1)\cr
b \ |n) \eq \sqrt{n+\nu} \ \ |n-1) \ \ \ for \ \ n \neq 0 \cr
b \ |0) \eq 0
\eea

If we now set
\bea
x_1 \eq \sqrt {\theta \over 2}  (b+b^{\dag}) \cr
x_2 \eq  \sqrt {\theta \over 2}  i(b-b^{\dag}).
\eea
we find that (10.4) is satisfied.

It is also possible to see that the solution is given in the
generalized Coulomb gauge as defined by
minimizing the matrix version of
(3.15).
\bea
\delta M \eq 0 \cr
M \eq Tr (x_i-y_i)^2.
\eea
In this equation the a variation of $x$ is defined by
\be
\delta x = i [\lambda,x]_m
\ee
where  $\lambda $ is an hermitian matrix.
The variational condition leads to the gauge condition
\be
\sum_{i=1,2} [x_i,y_i]_m =0.
\ee
{}From (10.2) and (10.7)
the condition may be written as
\be
[b,\adag]_m + [\bdag , a]_m =0.
\ee
This can easily be
confirmed from the defining properties of the operators.
Thus we have found an exact Coulomb gauge
classical solution of abelian \nc \ \cs \
theory at level 1. Multiple \qp \ solutions are easy to find but
we will not do so here.

 \setcounter{equation}{0}
\section{Conclusions }

In this paper we have reviewed the derivation of the \cs \
description of the \qh \ fluid (for $\nu =1/n$) given in \cite{safi}.
The
appropriate \cs \ theory has as its gauge invariance the group of
area preserving diffeomorphisms. In the linearized approximation
it becomes a conventional abelian \cs \ theory which efficiently
describes the large distance physics including the charge and
statistics of the quasiparticles. In a crude quantization one can
see qualitatively, but not quantitatively, the origin of the
quantization of the fluid density for the simplest filling
fractions.

In order to correctly capture the granular structure of the fluid
we upgraded the \cs \ theory to a \nc \ gauge  theory. The theory
can also be thought of as a matrix theory of the elementary
charges. The matrix theory is rich enough to describe fermions,
bosons or anyons in a strong magnetic field.  The \nc \ theory
exactly reproduces the quantitative connection between filling
fraction (level in the \cs \ description) and statistics required
by Laughlin's theory.

There are interesting predictions about the \nc \ \cs \ theory
that follow from the correspondence. An example is the phase
transition between \qh \ fluid behavior and the Wigner crystal
that occurs at low filling fraction. This suggests a phase
transition in the \nc \ \cs \ theory at large level.  The
transition would be associated with the spontaneous breaking of a
symmetry under area preserving diffeomorphisms of real $x$ space.
Similar phenomena have been observed fairly generically in \nc \
quantum field theories \cite{stripegubser}.

Finally in conclusion a speculation will be offered concerning the
generalization to more general filling factors for electrons. For
example consider the case $\nu = p/n$ with $p$ and $n$ relatively
prime. One way to try to construct such a state is to first
imagine $p$ non--interacting layers of \qh \ fluid, each with
filling fraction $1/n$. This can be represented in an obvious way
by a matrix theory of block diagonal matrices where the number of
blocks is $p$. We can think of the layers as branes separated by a
large enough distance so that the electrons can't tunnel between
them.

When the layers are adiabatically brought together so that the electrons

are easily shared between them, the state must approach the
fractional \qh \ state with $\nu = p/n$.
Experience with D-branes suggests that the resulting theory should
be a non--abelian version of the gauge theory. The natural guess
is \nc \ \cs \  $U(p)$ theory at level $n$.

\section{Acknowledgements }
I am very grateful to Nick Toumbas for extremely  helpful
discussions concerning the connections between \nc \ gauge theory,
the  Seiberg Witten map and its connection with the two coordinate
systems $x$ and $y$.  I would also like to thank M.M. Sheikh
Jabbari for valuable discussions and for correcting a number of
errors.  The author would also like to thank Ben Freivogel for
discussions.


\begin{thebibliography}{999}

\bibitem{dipole} M.M. Sheikh-Jabbari,
 Open Strings in a B-field Background as Electric Dipoles,
hep-th/9901080, Phys.Lett. B455 (1999) 129-134

\bibitem{bigatti} Daniela Bigatti, Leonard Susskind,
 Magnetic fields, branes and noncommutative geometry,
 hep-th/9908056,
 Phys.Rev. D62 (2000) 066004

\bibitem{safi} Safi Bahcall, Leonard Susskind,
Fluid Dynamics, Chern-Simons Theory and the Quantum Hall Effect,
Int.J.Mod.Phys.B5:2735-2750,1991

\bibitem{scz} By S.C. Zhang, T.H. Hansson, S. Kivelson,
An Effective Field Theory Model for the Fractional Quantum Hall
Effect, Phys.Rev.Lett.62:82-85,1988

 \bibitem{bob} Laughlin, R. B., 1987, in The Quantum Hall Effect,
edited by R. E. Prange and S. M. Girvin (Springer, Heidelberg), p.
233.





\bibitem{connes} A. Connes, Noncommutative Geometry, Academic
Press (1994)

\bibitem{rpm}  R.P. Manvelyan, R.L. Mkrtchyan,
Geometrical action for $w_\infty$ algebra as a reduced
symplectic Chern-Simons, hep-th/9401032,
Phys. Lett. B327 (1994) 47-49


\bibitem{bfss} T. Banks, W. Fischler, S.H. Shenker, L. Susskind,
 M Theory As A Matrix Model: A Conjecture,  hep-th/9610043,
Phys.Rev. D55 (1997) 5112-5128



\bibitem{t-nut}  B.A.Bernevig, J. Brodie, L. Susskind, N. Toumbas,
How Bob Laughlin Tamed the Giant Graviton from Taub-NUT space,
hep-th/0010105



\bibitem{kluson} J. Kluson,
 Matrix model and noncommutative Chern-Simons theory,
hep-th/0012184



\bibitem{poly}  Alexios P. Polychronakos,
 Noncommutative Chern-Simons terms and the noncommutative vacuum,
hep-th/0010264,  JHEP 0011 (2000) 008



\bibitem{seibwitt}  Nathan Seiberg, Edward Witten,
String Theory and Noncommutative Geometry,
 hep-th/9908142,
 JHEP 9909:032,1999,




\bibitem{seib} Nathan Seiberg,
A Note on Background Independence in Noncommutative Gauge
Theories,
  hep-th/0008013,  JHEP 0009 (2000) 003


\bibitem{gubser}  Steven S. Gubser, Mukund Rangamani,
 D-brane Dynamics and the Quantum Hall Effect
hep-th/0012155


\bibitem{stripegubser}  Steven S. Gubser, Shivaji L. Sondhi,
Phase structure of non-commutative scalar field theories,
 hep-th/0006119

\bibitem{solitons}  Rajesh Gopakumar, Shiraz Minwalla, Andrew
Strominger,  Noncommutative Solitons,
hep-th/0003160,  JHEP 0005 (2000) 020

See also Alexios P. Polychronakos,
 Flux tube solutions in noncommutative gauge theories,
  hep-th/0007043,
Phys.Lett. B495 (2000) 407-412



\end{thebibliography}
\end{document}